\documentclass[aps,superscriptaddress,twocolumn,twoside,floatfix,pra,a4paper,nofootinbib,groupedaddress]{revtex4-2} 

\usepackage{times}
\usepackage{epsfig}
\usepackage{amsfonts}
\usepackage{amsmath}
\usepackage{amssymb,amsthm}
\usepackage{xcolor,colortbl}
\usepackage{multirow}
\usepackage{braket}
\usepackage{latexsym}
\usepackage{amsfonts}
\usepackage{mathrsfs}
\usepackage{natbib}
\usepackage{verbatim}
\usepackage{gensymb}
\usepackage{caption}
\usepackage{caption}
\usepackage{subcaption}
\usepackage{ragged2e}
\usepackage{oplotsymbl}
\usepackage{array}
\usepackage{float}
\usepackage{lipsum}
\usepackage{diagbox}
\usepackage{tikz}

\usepackage{soul}
\DeclareCaptionJustification{justified}{\justifying}
\usepackage{blkarray}
 \usepackage{graphicx}

\usepackage[colorlinks=true,linkcolor=blue,citecolor=magenta,urlcolor=blue]{hyperref}
\allowdisplaybreaks

\begin{document}

\title{When Mei-Gu Guan's 1960 Postmen Get Empowered with Bell's 1964 Nonlocal Correlations\\
or\\ Nonlocal Advantage in Vehicle Routing Problem}

\author{Abhishek Banerjee} 
\affiliation{Centre for Astroparticle Physics and Space Science, Bose Institute,
EN 80, Sector V, Bidhannagar, Kolkata 700091, India.}

\author{Pratapaditya Bej}
\affiliation{Centre for Astroparticle Physics and Space Science, Bose Institute,
EN 80, Sector V, Bidhannagar, Kolkata 700091, India.}

\author{Amit Mukherjee}
\affiliation{Indian Institute of Technology  Jodhpur, Jodhpur 342030, India.}

\author{Sahil Gopalkrishna Naik}
\affiliation{Department of Physics of Complex Systems, S. N. Bose National Center for Basic Sciences, Block JD, Sector III, Salt Lake, Kolkata 700106, India.}

\author{Mir Alimuddin}
\affiliation{Department of Physics of Complex Systems, S. N. Bose National Center for Basic Sciences, Block JD, Sector III, Salt Lake, Kolkata 700106, India.}

\author{Manik Banik}
\affiliation{Department of Physics of Complex Systems, S. N. Bose National Center for Basic Sciences, Block JD, Sector III, Salt Lake, Kolkata 700106, India.}

\begin{abstract}
Vehicle routing problems, a comprehensive problem category originated from the seminal Chinese Postman Problem (first investigated by Chinese mathematician Mei-Gu Guan), entail strategic and tactical decision making for efficient scheduling and routing of vehicles. While Chinese postman problem is aimed at finding the minimum length cycle for a single postman, the broader challenges encompass scenarios with multiple postmen. Making cost-effective decisions in such cases depends on various factors, including vehicle sizes and types, vehicle usage time, road tax variations across routes, and more. In this work, we delve into a class of such problems wherein Bell nonlocal correlations provide advantages in optimizing the costs for non-communicating postmen, and thus establish a nascent utilization of quantum entanglement in traffic routing problem. Our investigation unveils promising applications for nonlocal correlations within combinatorial optimization and operational research problems, which otherwise have predominantly been explored within the quantum foundation and quantum information theory community.
\end{abstract}

\maketitle
{\it Introduction.--} The seminal 1964 theorem by John S. Bell drastically revolutionized our worldview \cite{Bell1964} (see also \cite{Bell1966, Mermin1993}). In an attempt to addressing a longstanding foundational debate \cite{Einstein1935, Bohr1935, Schrodinger1935}, Bell introduced an inequality whose violation, as demonstrated in several milestone experiments \cite{Freedman1972, Aspect1981, Aspect1982, Aspect1982(1), Zukowski1993} (see \cite{Wiki} for an comprehensive list of other important Bell tests), establishes the incompatibility of quantum theory with local hidden variable models. Subsequently, the advent of quantum information theory has facilitated practical utility to Bell inequalities by identifying numerous usages of quantum correlations violating such an inequality. Applications of quantum nonlocal correlations include establishing cryptographic keys between distant parties, certifying and expanding true randomness, witnessing the dimension of Hilbert spaces, providing efficient equilibrium strategies in Bayesian games, enhancing the zero-error capacity of noisy communication channels, and more \cite{Ekert1991, Barrett2005a, Acin2006, Pironio2010, Chaturvedi2015, Colbeck2012, Brunner2008, Gallego2010, Mukherjee2015, Cubitt2010, Cubitt2011, Alimuddin2023, Brunner2013, Pappa2015, Roy2016, Banik2019} (see also \cite{Brunner2014} and references therein).

Here, we report on a useful application of nonlocal correlations in real-world problems that have been extensively studied within the combinatorial optimization and operational research community. We delve into a category of problems known as Vehicle Routing Problems (VRPs) \cite{Toth2001}, which stand as an extension of the famous  Chinese Postman Problem (CPP) \cite{Kwan1960}. CPP was first explored by the mathematician Mei-Ko Kwan (modern PinYin spelling: Mei-Gu Guan) in 1960, amidst the Great Leap Forward era \cite{Grotschel2012}. Guan tackled an optimized variant of the renowned Euler’s bridges problem \cite{Euler1736}. While Euler inquired whether one can traverse a graph by passing through each edge exactly once, the CPP delves into determining the minimum aggregate length of additional edges required to render a graph Eulerian. These intricacies collectively fall under the umbrella of Arc Routing Problems \cite{Laporte2015}. Lately, there has been a burgeoning body of research on VRPs \cite{Toth2001}. In a VRP scenario, a fleet of vehicles is entrusted with a set of transportation requisites, prompting the quest for an optimal compilation of vehicle routes that fulfill these demands while economizing resources. The practical applications of VRPs are diverse, encompassing delivery logistics, transportation networks, and more (as exemplified by \cite{Tung2000,Park2010}). The appeal of these problems stems from their practical significance and the intricate nature of determining optimal solutions. This intricacy is further underscored by the classification of VRPs as NP-hard problem in general \cite{Toth2001}. 

In this work we demonstrate how nonlocal correlations can offer advantages in tackling intricate challenges of VRPs when multiple postmen /delivery personnel are involved. We consider the situation where distant postmen do not share communication among them and hence become ignorant about the choices of other fellows' delivery routes. In such a scenario, correlations provided to the postmen as assistance through the traffic signals can help them to choose their individual routes in a cost-efficient manner. Through concrete instances we establish how nonlocal correlations, acting as intermediaries between remote traffic signals, can substantially enhance the efficiency of complex vehicle routing problems. We explore the simplest Bell scenario that involves two distant parties each having two different inputs and each input yielding two different outputs -- a configuration succinctly termed as the $222$-Bell scenario \cite{Brunner2014}. The nonlocal nature of correlations observed in such experiments is depicted through the violation of different Bell-type inequalities. Within this framework, we consider two specific inequalities: the Clauser-Horne-Shimony-Holt (CHSH) inequality \cite{Clauser1969} and the tilted-CHSH inequality \cite{Acin2012}, and showcase the advantages of nonlocal correlations violating these inequalities in two distinct classes of simplistic VRPs. 
\begin{figure}[t!]
\includegraphics[width=0.45\textwidth, height=5cm]{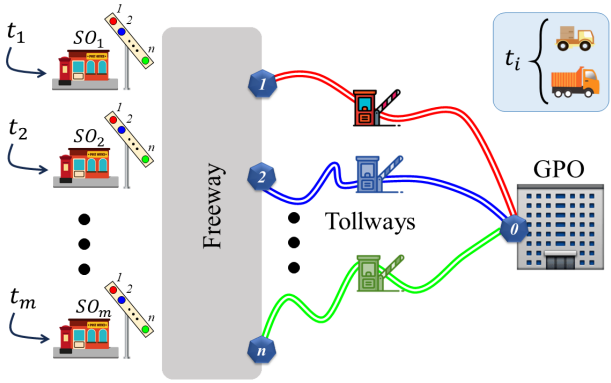}
\caption{An instance of VRP. One post-person from each of the $m$ distant SOs aims to deliver parcels to the GPO. The post-persons are provided different types of vehicle that they can drive through different tollways. Toll-costs across different tollways can vary depending on traffic-flow which further determine the incentive received by the post-persons who also gets additional rewards depending on how fast they can complete their trip. The post-persons can be assisted through correlated traffic signals to choose the tollways that maximize their total earnings.}\label{fig1}
\end{figure}

{\it The problem setup.--} Consider a network consisting of $m$ geographically distinct regional sub-post offices (SOs) entrusted with the responsibility of delivering parcels to a central hub -- the General Post Office (GPO). To facilitate this process, the postal department employs a courier service. During each delivery round, one post-person transports parcels from each SO to the GPO. The selection of the vehicle type allocated to the post-persons is contingent upon the size of the parcels and is drawn from a set ${1, 2, ..., k}$, which we represent as $[k]$. The choice of the vehicle type, denoted as $t_i$, for the $i^{th}$ post-person is made by the postal department and remains unknown to the post-person beforehand. The post-persons lack any means of communication amongst themselves. Consequently, the $i^{th}$ post-person is unaware of the vehicle types, $t_j$, assigned to their colleagues, where both $i$ and $j$ are within the range $[m]$, and $i\neq j$. There exist $n$ distinct tollways, each of which can be used by a post-person at their discretion to reach the GPO. The scenario can be depicted using a directed graph with nodes labeled as $v\in\{0, 1, \cdots, n\}\equiv [[n]]$. In this representation, node `$0$' corresponds to the GPO, while the other nodes correspond to the entry points of the tollways. Therefore, the directed edges $e_l:=(l, 0)$ represent various tollways, with $l\in [n]$ (see Figure \ref{fig1}). The flow of traffic on the $l^{th}$ tollway reflects the number of different types of vehicles traveling on that specific route. This traffic flow can be expressed as a $k$-tuple $\vec{N}_l=(n^1_l, n^2_l, \cdots, n^k_l)$.

The salary of each individual post-person for each delivery round is contingent upon the type of vehicle assigned to them. Precisely, the $i^{th}$ post-person operating a vehicle of type $t_i$ earns a salary denoted as $\$_i(t_i)$, where $t_i\in [k]$ for all $i\in [m]$. Toll costs for different vehicle types typically vary. Furthermore, the toll cost for a specific tollway may also be influenced by the number of distinct vehicle types passing through it during each delivery round. Different tollway operators, responsible for maintaining various routes, often offer attractive discount policies to attract more traffic to their respective paths. In addition to their base salary, post-persons receive incentives from their employer based on the discounts they can secure on the toll charges. These incentives, denoted as $\mathcal{I}_i$, are received by the $i^{th}$ post-person and depend on the traffic flow of the tollway they choose, {\it i.e.}, $\mathcal{I}_i \equiv \mathcal{I}_i(l,\vec{N}_l)$. Post-persons also strive to minimize their travel times to maximize their rewards ($\mathcal{R}$) offered by the courier service for completing a greater number of deliveries. This motivates them to avoid congested tollways. The travel time of the $i^{th}$ post-person, and consequently their corresponding reward $\mathcal{R}_i$, is influenced by the traffic flow, {\it i.e.}, $\mathcal{R}_i \equiv\mathcal{R}_i(l,\vec{N}_l)$. In addition to fulfilling their delivery obligations, the post-persons endeavor to optimize their collective earnings by making prudent decisions regarding their delivery routes. The choices of delivery routes made by the post-persons define a `path configuration' within the aforementioned traffic network. Therefore, from a mathematical perspective, the collective objective of the post-persons can be formulated as the following optimization problem:
\begin{align}
&~~~~~~~\mathcal{E}^\star=\max_{\text{path configurations}}\mathcal{E}_{PC},\label{earning}\\
\mathcal{E}_{PC}&:=\sum_{i=1}^m\left[\$_i(t_i)+\mathcal{I}_i(l,\vec{N}_l)+\mathcal{R}_i(l,\vec{N}_l)\right],\nonumber
\end{align}
where $\mathcal{E}_{PC}$ denote total earning in a particular path configuration and maximization is done over all possible path configurations. In the above, while we have considered that the incentive and reward depend only on the traffic flow and the tollway, and the salary depends on the type of vehicles only, in real-world problems several other factors might play crucial role. For instance, the parcels could be categorized depending on how urgently they need to be delivered, and there could be no toll cost on certain parcels, such as  the medical related items.    

{\it Assistance through traffic signal.--} In order to choose a path that eventually maximizes their collective earnings the post-persons can follow different strategies. In a deterministic strategy each post-persons chooses a definite tollway based on the vehicle type they are assigned. Mathematically a deterministic strategy can be represented as a mapping from the set of vehicle types to the tollways, {\it i.e.}, $f:t \mapsto l$. One can consider a more general mixed strategy wherein depending upon the vehicle types different tollways are chosen in a stochastic manner, {\it i.e.}, $i^{th}$ post-person's strategy is a probability distribution $\{p(l_i|t_i)~|~p(l_i|t_i)\ge0~\&~\sum_{l_i=1}^np(l_i|t_i)=1,~\forall~t_i\in[k],~i\in[m]\}$. Operationally, this can be achieved through traffic poles at the SOs' that instruct the post-persons to choose their tollways (see Fig. \ref{fig1}). The combined strategy is thus a product probability distribution $p(~\vec{l}~|~\vec{t}~):=\Pi_{i=1}^mp(l_i|t_i)$, where $\vec{l}\in[n]^{\times m}$ and $\vec{t}\in[k]^{\times m}$. The post-persons can follow even a more general correlated strategy wherein the probability distribution $p(~\vec{l}~|~\vec{t}~)$ is not factorized in the above form. In other words, the strategy of different post-persons are not independent anymore, rather their actions become correlated. At this point we will assume the correlation $p(~\vec{l}~|~\vec{t}~)$ to satisfy the no-signaling (NS) conditions, that prohibits any information transfers among the post-persons to known each other vehicle types. Assistance of such correlation is vastly studied in Bayesian game theory where the players have incomplete information \cite{Harsanyi1968}. The solution concept of correlated equilibrium -- a general notion than the well known concept of Nash equilibrium \cite{Nash1950} -- turns out to be natural while studying the Bayesian games \cite{Aumann1974,Aumann1987,Papadimitriou2008}. Here we aim to study how such correlations provided as assistance through traffic signals can be advantageous in optimizing the earnings.

{\it Local vs Nonlocal Assistance.--} Interestingly, the correlations $p(~\vec{l}~|~\vec{t}~)$ can be of two type -- classical and beyond. J. S. Bell in his seminal work \cite{Bell1964} (see also \cite{Bell1966}) have shown that any correlation obtained in classical world allows a {\it local-causal} description. Mathematically this boils down to the fact that any classical correlation can be expressed as $p(~\vec{l}~|~\vec{t}~)=\int_\Lambda d\lambda~p(\lambda)\Pi_{i=1}^mp(l_i|t_i,\lambda)$, where $\lambda\in\Lambda$ is some classical variable (popularly called `hidden variable' within quantum foundation community) shared among the distant traffic signals, and $p(\lambda)$ in a probability distribution on $\Lambda$. Correlations that are not local are called Bell-nonlocal or in short nonlocal. Often, nonlocal nature of a correlation is established through violation of some Bell type inequalities \cite{Brunner2014}. Interestingly, multipartite quantum systems prepared in entangled states lead to violations of such inequalities, and thus establishes the nonlocal nature of quantum world \cite{Freedman1972, Aspect1981, Aspect1982, Aspect1982(1), Zukowski1993}.

For instance, in the scenario involving two parties, each performing two dichotomic measurements (the $222$ scenario), the joint probability distribution can be expressed as 
\begin{align}\nonumber
&(p(11|ij),p(12|ij),p(21|ij),p(22|ij))\\\label{ns}
&~~~~~~~\equiv (c_{ij},m_{i}-c_{ij},n_{j}-c_{ij},1-m_{i}-n_{j}+c_{ij}),  
\end{align}
where $m_{i}$ and $n_{j}$ respectively denote the marginal probabilities of outcome `1' for Alice's $i^{th}$ input and Bob's $j^{th}$ input. The Bell CHSH expression for this correlation reads as 
\begin{align}
\mathbb{B}=2+4\left(c_{00}+c_{01}+c_{10}-c_{11}\right)-4\left(m_0+n_0\right),\label{bell}
\end{align}
where, $\mathbb{B}:=\langle 11\rangle+\langle 12\rangle+\langle 21\rangle-\langle 22\rangle$, with $\langle ij\rangle:=\sum_{a,b=1}^2(-1)^{a+b}p(ab|ij)$. A $222$ correlation will allow a local-causal description {\it if and only if} it satisfies the Bell-CHSH inequality $|\mathbb{B}|\le2$ and its relabeled versions \cite{Fine1982}. Interestingly, bipartite quantum systems prepared in entangled states can lead to correlations that violate this inequality and thus become nonlocal. In the next, we will demonstrate with explicit examples how such nonlocal correlations prove advantageous in optimizing the quantities in Eq.(\ref{earning}).   

{\it An Example.--} Consider a VRP that consists of two SOs each connected to the GPO through two different tollways, and post-persons are assigned two types of vehicles, either a small one or a large one, {\it i.e.}, $m,l_i,t_i\in[2]$ for $i\in[2]$. The salary, the incentive, and the reward provided to post-persons depends on the type of vehicles assigned and the tollways they travel (see the payoff matrix in Table \ref{payoff}). 
\begin{table}[t!]
\centering
\begin{tabular}{cc|cc|}
\cline{3-4}
&      & \multicolumn{2}{c|}{$l_1l_2$}\\ \cline{3-4} 
&      & \multicolumn{1}{c|}{$11, 22$}         & $12, 21$\\ \hline
\multicolumn{1}{|c|}{\multirow{4}{*}{$t_1t_2$}} & $~~11~~$ & \multicolumn{1}{c|}{$(2s,u_1, x_1)$} & $(2s,z_1,y_1)$ \\ \cline{2-4} 
\multicolumn{1}{|c|}{} & $12$ & \multicolumn{1}{c|}{$~~~~~(s+l,u_2, x_2)~~~~~$} & $~~~~~(s+l,z_2,y_2)~~~~~$ \\ \cline{2-4} 
\multicolumn{1}{|c|}{} & $21$ & \multicolumn{1}{c|}{$(s+l, u_3,x_3)$} & $(s+l,z_3,y_3)$ \\ \cline{2-4} 
\multicolumn{1}{|c|}{} & $22$ & \multicolumn{1}{c|}{$(2l,u_4,x_4)$}   & $(2l,z_4,y_4)$  \\ \hline
\end{tabular}
\caption{The Payoff Matrix. The payoffs corresponding to the total salary, incentive and reward are provided in parentheses and determined by $18$ real parameters. For instance, if the assigned vehicles are $t_1=1$, $t_2=1$ and the chosen tollways are $l_1=1$, $l_2=1$ the respective payoffs are given by $(2s,u_1,x_1)$.} \label{payoff}
\end{table}
The payoff elements are chosen following real-world rational requirements. 
\begin{itemize}
\item[(i)] The salary $s$ for driving the small vehicle is assumed to be less than the salary $l$ of the large vehicle, {\it i.e.}, $s<l$.
\item[(ii)] Different operators operating the tollways offer discounts over the toll cost to attract more traffic. The discount depends on the number as well as the size of the vehicles. This discount amount is provided to the post-persons as incentive. In Table \ref{payoff} we assume $u_j\le u_{j+1}~\&~z_{j+1}\le z_j$ and $z_1\le u_1$.
\item[(iii)] The post-persons are provided more rewards depending on how fast they complete the trip. Accordingly, $\mathcal{R}_i$'s are inversely proportional to tollway congestion and directly proportional to vehicle speed. It is natural to assume that vehicle speed is inversely proportional to vehicle size. Thus in Table \ref{payoff} we have $x_{j+1}\le x_j~\&~y_{j+1}\le y_{j}$. Furthermore we have $x_2=x_3:=x$ and $y_2=y_3:=y$ which tells that rewards do not change under swapping of the paths. Additionally, we have $x_1\le y_4$ denoting the fact that vehicles moving in different paths face no congestion at all. 
\end{itemize}
While the requirements (i)-(iii) are fixed in a particular way and represent a broad class of VRPs, one can fix the requirement in other ways too. We now proceed to analyze a smaller set of VRPs in the above framework by fixing 
\begin{align}\label{vrpr6}
\left\{\!\begin{aligned}
u_1=2u_s,~u_4=2u_l,~u_2=u_3=u_s+u_l,\\  
x_1=x+(l-s)+(u_l-u_s),~{y< u_l+u_s+x},\\
x_4=y-(l-s)-2u_l,~z_1=z_2=z_3=z_4=0,\\
y_1=y+(l-s),~y_4=x-(l-s)+(u_s+u_l),
\end{aligned}\right\};
\end{align}
where $u_s$ and $u_l$ refer to the incentive corresponding to small and large vehicle, respectively. Note that, the constraint $x_1\le y_4$ implies $(l-s)\le u_s$. The VRP in Eq.(\ref{vrpr6}) is specified by six independent variables $s, l, u_l, u_s, x, y$. Considering all these constraints, a set of feasible VRPs exists within a six-dimensional region specified by:
\begin{align}\label{sol}
\left\{\!\begin{aligned}
0<s<l, ~~  l - s \leq u_s < u_l ~\\~ 0< x < y < u_l + u_s + x
\end{aligned}\right\}.
\end{align}
As already pointed out, the post-persons aim to maximize their total earnings. Please note that, in Eq.(\ref{earning}) the optimization is formulated under the consideration that the post-persons follows a definite path configuration. However, given a NS correlation (\ref{ns}) they can follow more general strategies by choosing different path configurations in correlated manners.  Given such a correlation as assistance through traffic signal their total earning read as   
\vspace{-.15cm}
\begin{align}
\mathcal{E}_{\mathbb{P}}:=\mbox{Tr}[(\sum_i \left(\$_i+ \mathcal{I}_i+\mathcal{R}_i\right))^{\mathrm{T}}\cdot \mathbb{P}],
\end{align}
where the elements of $\mathbb{P}$ are given by  $p(l_1l_2t_1t_2):=p(t_1t_2)p(l_1l_2|t_1t_2)$, and $\mathrm{T}$ denotes transpose of the resulting payoff matrix obtained from Table \ref{payoff}. For the case $p(t_1t_2)=\frac{1}{4}$, the earning of the VRP (\ref{vrpr6}) becomes (see Appendix A) 
\begin{align*}
\mathcal{E}_{\mathbb{P}}:=\left(l + s + \frac{u_{l}+u_{s}+x+y}{2}\right) +
\frac{\left(u_{l} + u_{s} + x - y\right) \mathbb{B}}{8}, 
\end{align*}
an increasing linear function of the Bell CHSH expression.
The highest value $\mathbb{B}$ can achieve classically is $2$, but, by using a quantum state $\ket{\psi}_{AB}=\left( \ket{00}_{AB}+\ket{11}_{AB}\right)/\sqrt{2}$ with the local projective measurements $M_{t_1=1}:=\sigma_z$, $M_{t_1=2}:=\sigma_x$ and $M_{t_2=1}:=\left({\sigma_z+\sigma_x}\right)/{\sqrt{2}}$, $M_{t_2=2}:=\left({\sigma_z-\sigma_x}\right)/{\sqrt{2}}$, $\mathbb{B}$ can reach the Tsirelson's bound of $2\sqrt{2}$ . This results in a higher expected payoff $\mathcal{E}_{\mathbb{P}}$ in the quantum case than in the classical case. A natural question arises: are there VRP games where non-maximally entangled states are the best resource states? To that end, we propose another class of VRPs by increasing the incentives for vehicle allotment of types $t_1$ and $t_2$ and by adding a positive quantity $2\zeta$ to the payoffs for choosing routes 11 and 12, respectively. All other payoffs, conditions, and constraints remain the same (see Appendix B). This makes the new average earning $\mathcal{E'}_{\mathbb{P}}:=\left(l + s + \frac{u_{l}}{2} + \frac{u_{s}}{2} + \frac{x}{2} + \frac{y}{2} \right) +\frac{1}{8}{\left(u_{l} + u_{s} + x - y\right) (\mathbb{B}+2\zeta m_0)}$. Note that, $(\mathbb{B}+2\zeta m_0)$ is tilted Bell type expression \cite{Acin2012}. The optimum value of this expression in quantum regime can be obtained using a quantum state $\ket{\psi}_{AB}=\cos\theta\ket{00}_{AB}+\sin\theta\ket{11}_{AB}$ with the local projective measurements $M_{t_1=1}:=\sigma_z$, $M_{t_1=2}:=\sigma_x$ and $M_{t_2=1}:=\cos\beta\sigma_z+\sin\beta\sigma_x$, $M_{t_2=2}:=\cos\beta\sigma_z-\sin\beta\sigma_x$; where $\tan\beta=\sin 2\theta$ and $\zeta=2/\sqrt{1+2\tan^22\theta}$ \cite{Yang2013, Bamps2015}.

\textit{Discussion.--} Throughout the history of scientific discovery, disciplines with no apparent common ground, and at times belonging to different scientific spheres, have revealed unsuspected connections. Occasionally, these intersections have yielded remarkable and highly fruitful outcomes. For instance, while Einstein's formulation of General Relativity in 1915 utilized the mathematical tools of differential geometry originally introduced by Riemann in 1854, Noam Chomsky's significant work on linguistics influenced the development of formal language theory in computer science. With a similar spirit, in this work, we establish an intriguing connection between quantum foundations and operational research problems. Particularly, we demonstrate utility of quantum nonlocal correlations in combinatorial optimization and operational research problems. The nonlocal behavior of the quantum world was first pointed out by J. S. Bell in his seminal 1964 work, whereas vehicle routing problems emerge as generalizations of the famous Chinese Postman Problem originally studied by Mei-Gu Guan in 1960, which subsequently gave birth to the field of combinatorial optimization and operational research. Despite these concepts being contemporary, the potential use of nonlocal correlations in those real-world optimization problems was missing until now.  

The present work extends an invitation to the community engaged in the exploration of nonlocal correlations, as well as to researchers dedicated to operational research, encouraging them to delve into more exotic and transformative applications of quantum nonlocal correlations in real-world problem-solving. At this juncture, we draw attention to recent and pertinent contributions in the realm of network communication theory \cite{Quek2017,Leditzky2020, Yun2020}. In these studies, authors have adeptly showcased how distant senders, united in their goal to communicate with a common receiver, can gain advantages by correlating their encoding strategies through utilization of nonlocal correlations. While communication setups strive to optimize simultaneously achievable capacity regions, the focus in VRPs shifts towards maximizing the overall payoff for the involved parties. The intricacies in the VRPs we examine stem from the delicate balance between payoffs as incentives and rewards. The escalation of congestion results in diminished rewards but heightened incentives, transforming the problem into a manifestation of Multi-Criteria Decision Making (MCDM) challenges \cite{book}. This particular problem class has garnered substantial attention in the domain of Operations Research, with examples ranging from the intricacies of portfolio management --meticulously selecting optimal financial instruments from a dynamic market landscape to the subtleties of personnel selection, where decision-makers endeavour to identify the most suitable candidates amid a myriad of criteria, energy planning etc. \cite{MCDMPPM, Personnel, MCDMenergy}. While our investigation sheds light on the quantum advantage within a specific family of VRP scenarios, a broader goal is to find quantum advantages in other classes of MCDM problems.

\begin{acknowledgements}
SGN acknowledges support from the CSIR project 09/0575(15951)/2022-EMR-I. MA and MB acknowledge funding from the National Mission in Interdisciplinary Cyber-Physical systems from the Department of Science and Technology through the I-HUB Quantum Technology Foundation (Grant no: I-HUB/PDF/2021-22/008). MB acknowledges support through the research grant of INSPIRE Faculty fellowship from the Department of Science and Technology, Government of India, and the start-up research grant from SERB, Department of Science and Technology (Grant no: SRG/2021/000267).
\end{acknowledgements}

\onecolumngrid
\section{Appendix A: CHSH inequality and a class of VRPs} \label{appen-a}
We begin by presenting the complete payoff matrix of the Table \ref{payoff}.
\begin{table}[h!]
 \begin{tabular}{c|c|c|c|c|}
      \diagbox{$t_1,t_2$}{$l_1,l_2$} & 11 & 12 & 21 & 22 \\
      \hline
      11&  $(2s, u_1, x_1)$ & $(2s, z_1,  y_1)$ & $(2s, z_1, y_1)$ & $(2s, u_1, x_1)$\\
      \hline
     12 & $(s + l, u_2, x_2)$ & $(s +l, z_2, y_2)$ & $(s+l, z_2, y_2)$ & $(s+l, u_2, x_2)$\\
      \hline
     21 & $(s + l, u_3, x_3)$ & $(s +l, z_3, y_3)$ & $(s+l, z_3, y_3)$ & $(s+l, u_3, x_3)$\\
      \hline
     22 & $(2l, u_4, x_4)$ & $(2l, z_4, y_4)$ & $(2l, z_4, y_4)$ & $(2l, u_4, x_4)$ \\
     \hline
 \end{tabular}
\end{table}
The payoffs corresponding to the total salary, incentive and reward are provided in parentheses and determined by $18$ real parameters. The total payoffs $\left(\sum_i \left(\$_i+ \mathcal{I}_i+\mathcal{R}_i\right)\right)$ taking into account the salary, the incentive, and the reward corresponding to different choices of vehicles $t_i \in \{1, 2\}$ and routes $l_i \in \{1, 2\}$ from Table \ref{payoff} can be represented by the following matrix.
\begin{table}[h!]
 \begin{tabular}{c|c|c|c|c|}
      \diagbox{$t_1,t_2$}{$l_1,l_2$} & 11 & 12 & 21 & 22 \\
      \hline
      11&  $(2s + u_1 + x_1)$ & $(2s + z_1 +  y_1)$ & $(2s + z_1 + y_1)$ & $(2s + u_1 + x_1)$\\
      \hline
     12 & $(s + l + u_2 + x_2)$ & $(s +l + z_2 + y_2)$ & $(s+l + z_2 + y_2)$ & $(s+l + u_2 + x_2)$\\
      \hline
     21 & $(s + l + u_3 + x_3)$ & $(s +l + z_3 + y_3)$ & $(s+l + z_3+ y_3)$ & $(s+l+ u_3+ x_3)$\\
      \hline
     22 & $(2l+ u_4 + x_4)$ & $(2l + z_4 + y_4)$ & $(2l + z_4 + y_4)$ & $(2l + u_4 + x_4)$ \\
     \hline
 \end{tabular}
\end{table}
Each element of the matrix below represents the sum of three payoffs: the salary, the incentive, and the reward corresponding to the same element in Table \ref{payoff}.

The considered VRP game is subject to the following requirements: (i) - (iii) and Eq.(\ref{vrpr6}). Straightforward algebraic manipulations reduce these constraints to the following (Eq. (\ref{sol})): 
\begin{align*}
0<s<l,~~ l - s \leq u_s < u_l,~~ 0< x < y < u_l + u_s + x .
\end{align*}
Besides, in the scenario involving two parties, each performing two dichotomic measurements (the $222$ scenario), the joint probability distribution that adheres to no-signaling conditions $\sum_{l_1}p(l_1l_2|t_1t_2)=\sum_{l_1}p(l_1l_2|t'_1t_2)~~ \forall t_1,t'_1,l_2,t_2$ and $\sum_{l_2}p(l_1l_2|t_1t_2)=\sum_{l_2}p(l_1l_2|t_1t'_2)~~ \forall t_2,t'_2,l_1,t_1$ is given by the following matrix as given in the Eq.(\ref{ns}):
\begin{table}[h!]
\begin{tabular}{c|c|c|c|c|}
\diagbox{$t_1, t_2$}{$l_1, l_2$} & 11 & 12 & 21 & 22 \\
\hline
11 & $c_{00}$ & $- c_{00} + m_{0}$ & $- c_{00} + n_{0}$ & $c_{00} - m_{0} - n_{0} + 1$\\
\hline
12 & $c_{01}$ & $- c_{01} + m_{0}$ & $- c_{01} + n_{1}$ & $c_{01} - m_{0} - n_{1} + 1$ \\
\hline
21 & $c_{10}$ & $- c_{10} + m_{1}$ & $- c_{10} + n_{0}$ & $c_{10} - m_{1} - n_{0} + 1$ \\
\hline
22 & $c_{11}$ & $- c_{11} + m_{1}$ & $- c_{11} + n_{1}$ & $c_{11} - m_{1} - n_{1} + 1$\\
\hline
\end{tabular}
\end{table}
If the traffic signals share this correlation as assistance, the players' total earnings are given by the average payoff expression:
\begin{align}
\mathcal{E}_{\mathbb{P}}:=\mbox{Tr}\left[\left(\sum_i \left(\$_i+ \mathcal{I}_i+\mathcal{R}_i\right)\right)^{\mathrm{T}}\cdot \mathbb{P}\right].
\end{align}
In this equation, the elements of $\mathbb{P}$ are defined as $p(l_1l_2t_1t_2) := p(l_1l_2|t_1t_2) {p(t_1t_2)}$, where $\mathrm{T}$ represents the transpose of the payoff matrix derived from Table \ref{payoff}. We assume that the choice of vehicles for each player is entirely random and independent of each other, indicating $p(t_1t_2)=p(t_1)p(t_2)=\frac{1}{4}$ for all $t_1,t_2\in \{1,2\}$. The average payoff, $\mathcal{E}_{\mathbb{P}}$ is thus given by,
\footnotesize
\begin{align*}
\mathcal{E}_{\mathbb{P}}&={\frac{1}{4}}\cdot\text{Tr}\left(\begin{pmatrix}
2s + u_1 + x_1 & 2s + z_1 +  y_1 & 2s + u_1 + x_1 & 2s + u_1 + x_1 \\
s + l + u_2 + x_2 & s +l + z_2 + y_2 & s+l + z_2 + y_2 & s+l + u_2 + x_2 \\
s + l + u_3 + x_3 & s +l + z_3 + y_3 & s+l + z_3+ y_3 & s+l+ u_3+ x_3 \\
2l+ u_4 + x_4 & 2l + z_4 + y_4 & 2l + z_4 + y_4 & 2l + u_4 + x_4
\end{pmatrix}^{\mathrm{T}}\cdot\begin{pmatrix}
c_{00} & -c_{00} + m_{0} & -c_{00} + n_{0} & c_{00} - m_{0} - n_{0} + 1\\
c_{01} & -c_{01} + m_{0} & -c_{01} + n_{1} & c_{01} - m_{0} - n_{1} + 1\\
c_{10} & -c_{10} + m_{1} & -c_{10} + n_{0} & c_{10} - m_{1} - n_{0} + 1\\
c_{11} & -c_{11} + m_{1} & -c_{11} + n_{1} & c_{11} - m_{1} - n_{1} + 1
\end{pmatrix}\right),\\
&=\left(l + s + \frac{u_{l}}{2} + \frac{u_{s}}{2} + \frac{x}{2} + \frac{y}{2} \right) + \frac{\left(u_{l} + u_{s} + x - y\right)}{8}\mathbb{B},
\end{align*}
\normalsize
where, $\mathbb{B}$ the CHSH expression (see Eq.(\ref{bell}) in the main manuscript).

\section{Appendix B: Tilted-CHSH inequality and a class of VRPs}\label{appen-b}

In this new game, we consider a somewhat modified scenario. 
{If both post persons are provided small cars, they will have the opportunity to transport some extra utensils. However, there are specific conditions to consider. Person $1$ must deliver the items to a designated location on route $1$, whereas person $2$ can choose to deliver them on either route $1$ or route $2$. Successful completion of this task will result in both individuals receiving an additional cash bonus of $\zeta$.  }Note that the only change is in the introduction of the constant $2\zeta$. The rest of the conditions all remain exactly the same as before.
The modified payoff table corresponding to this is presented below: 
\begin{table}[h!]
\begin{tabular}{|c|c|c|c|c|}
      \hline 
      \diagbox{$t_1,t_2$}{$l_1,l_2$} & 11 & 12 & 21 & 22 \\
      \hline
      11&  $(2s, u_1 + 2 \zeta, x_1)$ & $(2s, z_1 + 2 \zeta, y_1)$ & $(2s, u_1, x_1)$ & $(2s, u_1, x_1)$\\
      \hline
     12 & $(s + l, u_2, x_2)$ & $(s +l, z_2, y_2)$ & $(s+l, z_2, y_2)$ & $(s+l, u_2, x_2)$\\
      \hline
     21 & $(s + l, u_3, x_3)$ & $(s +l, z_3, y_3)$ & $(s+l, z_3, y_)$ & $(s+l, u_3, x_3)$\\
      \hline
     22 & $(2l, u_4, x_4)$ & $(2l, z_4, y_4)$ & $(2l, z_4, y_4)$ & $(2l, u_4, x_4)$ \\
     \hline
 \end{tabular}
\end{table}
The payoffs corresponding to the total salary, incentive and reward are provided in parentheses and determined by $19$ real parameters.

Similarly as in \ref{appen-a}, The total payoffs $\left(\sum_i \left(\$_i+ \mathcal{I}_i+\mathcal{R}_i\right)\right)$ taking into account the salary, the incentive, and the reward corresponding to different choices of vehicles $t_i \in \{1, 2\}$ and routes $l_i \in \{1, 2\}$ from the above table can be represented by the following matrix.
\begin{table}[h!]
 \begin{tabular}{c|c|c|c|c|}
      \diagbox{$t_1,t_2$}{$l_1,l_2$} & 11 & 12 & 21 & 22 \\
      \hline
      11&  $(2s + u_1 + x_1 + 2\zeta)$ & $(2s + z_1 +  y_1 + 2\zeta)$ & $(2s + u_1 + x_1)$ & $(2s + u_1 + x_1)$\\
      \hline
     12 & $(s + l + u_2 + x_2)$ & $(s +l + z_2 + y_2)$ & $(s+l + z_2 + y_2)$ & $(s+l + u_2 + x_2)$\\
      \hline
     21 & $(s + l + u_3 + x_3)$ & $(s +l + z_3 + y_3)$ & $(s+l + z_3+ y_3)$ & $(s+l+ u_3+ x_3)$\\
      \hline
     22 & $(2l+ u_4 + x_4)$ & $(2l + z_4 + y_4)$ & $(2l + z_4 + y_4)$ & $(2l + u_4 + x_4)$ \\
     \hline
     
 \end{tabular} 
\end{table}

Each element of the matrix above represents the sum of three payoffs: the salary, the incentive, and the reward corresponding to the same element in the above table.
The players' total earnings are given by the average payoff expression:
\begin{align}
\mathcal{E'}_{\mathbb{P}}:=\mbox{Tr}\left[\left(\sum_i \left(\$_i+ \mathcal{I}_i+\mathcal{R}_i\right)\right)^{\mathrm{T}}\cdot \mathbb{P}\right].
\end{align}
Where the payoff matrix is now the above table instead of the previous one.
Thus the average payoff now becomes
\footnotesize
\begin{align} 
\frac{1}{4}\cdot \text{Tr}\left(\scriptscriptstyle
\begin{pmatrix}
2s + u_1 + x_1 +2\zeta & 2s + z_1 +  y_1 +2\zeta & 2s + u_1 + x_1 & 2s + u_1 + x_1 \\
s + l + u_2 + x_2 & s +l + z_2 + y_2 & s+l + z_2 + y_2 & s+l + u_2 + x_2 \\
s + l + u_3 + x_3 & s +l + z_3 + y_3 & s+l + z_3+ y_3 & s+l+ u_3+ x_3 \\
2l+ u_4 + x_4 & 2l + z_4 + y_4 & 2l + z_4 + y_4 & 2l + u_4 + x_4
\end{pmatrix}^{\mathrm{T}}\cdot\begin{pmatrix}
c_{00} & -c_{00} + m_{0} & -c_{00} + n_{0} & c_{00} - m_{0} - n_{0} + 1\\
c_{01} & -c_{01} + m_{0} & -c_{01} + n_{1} & c_{01} - m_{0} - n_{1} + 1\\
c_{10} & -c_{10} + m_{1} & -c_{10} + n_{0} & c_{10} - m_{1} - n_{0} + 1\\
c_{11} & -c_{11} + m_{1} & -c_{11} + n_{1} & c_{11} - m_{1} - n_{1} + 1
\end{pmatrix} \right)
\end{align}
\normalsize
Which can be simplified to :

 $$\mathcal{E'}_\mathbb{P} = \left(l + s + \frac{u_{l}}{2} + \frac{u_{s}}{2} + \frac{x}{2} + \frac{y}{2} \right) + \frac{\left(u_{l} + u_{s} + x - y\right) (\mathbb{B} + 2\zeta m_0)}{8}$$

 Where $\mathbb{B}$ is given by Eq.(\ref{bell}).
The expression $\mathbb{B} + 2\zeta m_0$ is the so called Tilted-Bell expression. This expression attains a maximum value of $2\sqrt{2}$ for a particular choice of measurements and a particular non-maximally entangled state for each value of $\zeta$.

The rest of the arguments remain exactly the same as of Appendix \ref{appen-a}.

\twocolumngrid

\end{document}